\documentclass{article}
\usepackage{spconf,amsmath,graphicx}

\newcommand{\tabref}[1]{\mbox{Table~\ref{#1}}}
\newcommand{\figref}[1]{\mbox{Figure~\ref{#1}}}
\usepackage{url}
\usepackage{color}
\usepackage[english]{babel}
\usepackage[T1]{fontenc}
\usepackage[utf8]{inputenc}
\usepackage{booktabs} 
\usepackage{algorithm}
\usepackage{algorithmic}

\newcommand{\algref}[1]{\mbox{Algorithm~\ref{#1}}}
 
\title{Learning a Representation for Cover Song Identification Using Convolutional Neural Network}
\name{Zhesong Yu, Xiaoshuo Xu, Xiaoou Chen, Deshun Yang}
\address{Wangxuan Institute of Computer Technology, Peking University}
\begin{document}
%
\maketitle
\begin{abstract}
Cover song identification represents a challenging task in the field of Music Information Retrieval (MIR) due to complex musical variations between query tracks and cover versions. Previous works typically utilize hand-crafted features and alignment algorithms for the task. More recently, further breakthroughs are achieved employing neural network approaches. In this paper, we propose a novel Convolutional Neural Network (CNN) architecture based on the characteristics of the cover song task. We first train the network through classification strategies; the network is then used to extract music representation for cover song identification. A scheme is designed to train robust models against tempo changes. Experimental results show that our approach outperforms state-of-the-art methods on all public datasets, improving the performance especially on the large dataset. 
\end{abstract}
\begin{keywords}
Music Information Retrieval, Cover Song Identification
\end{keywords}

\section{Introduction}
\label{sec:intro}

Cover song identification has long been a popular task in the music information retrieval community, with potential applications in areas such as music license management, music retrieval, and music recommendation. Cover song identification can also be seen as measuring the similarity between music melodies. Given those cover songs may differ from the original song in key transposition, speed change and structural variations, identifying cover songs is a rather challenging task. Over the past ten years, researchers initially attempt to address the problem employing Dynamic Programming (DP) approaches. Typically, chroma sequences representing the intensity of twelve pitch classes are used to describe recordings, and then a DP method is utilized for finding an optimal alignment between two given recordings, resolving the discrepancy caused by tempo changes and structural variations \cite{ellis2007identifyingcover,serra2008chroma,serra2009cross}. 
Such approaches work well when facing structural variations and tempo changes in the music; however, the involvement of element-to-element distance computation with quadratic time complexity makes it unsuitable for large-scale datasets. 

Alternatively, some researchers attempted to identify cover songs by modeling the music. For instance, Serrà et al. studied time series modeling for cover song identification \cite{serra2009cross}. 
\cite{bertin2012large,osmalsky2016enhancing} represent the music with fixed-dimensional vectors, which enables a direct measure of the music similarity. 
These approaches highly improved the efficiency compared to alignment methods, while the loss of the temporal information of music in these approaches may yield a lower precision.

Moreover, deep learning approaches are introduced to cover song identification. For instance, CNNs are utilized to measure the similarity matrix \cite{chang_audio_2017} or learn features \cite{qi2017audio,xu2018key,yu2019temporal,DBLP:journals/corr/abs-1907-01824}. While these methods have achieved promising results, there is still room for improvement.
In this paper, a specially designed CNN architecture is proposed to overcome challenges of key transposition, speed change and structural variations existing in cover song identification. Notably, the use of the specialized kernel size is fist ever utilized in the field of music information retrieval. The dilation convolution and the method of data augmentation are also introduced. Our approach outperforms state-of-the-art methods on all public datasets with better accuracy but lower time complexity. Furthermore, to our best knowledge, our method is currently the best method to identify cover songs in huge real-life corpora.

\section{Approach}

\subsection{Problem Formulation}
\begin{figure}
	\centering
	\includegraphics[width=3.2in]{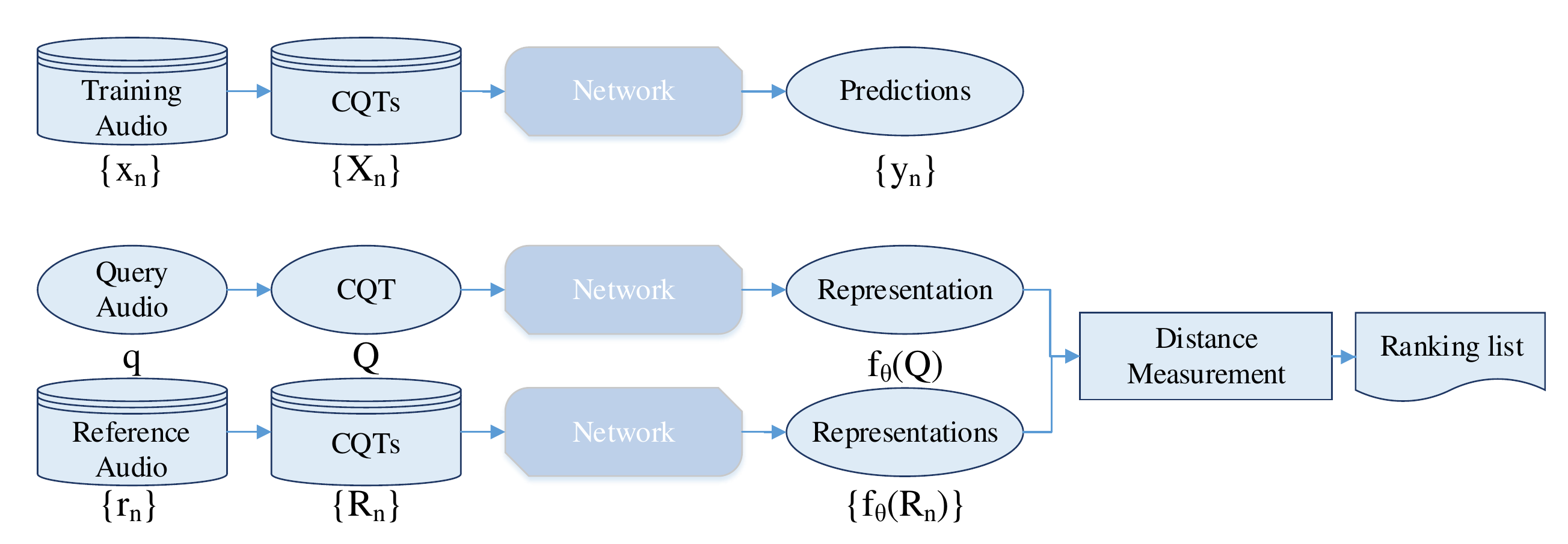}
	\caption{Training procedure and retrieval procedure.}
	\label{fig:pipe}
\end{figure}
As shown in \figref{fig:pipe}, we have a training dataset $D = \{(x_n, t_n)\}$, where $x_n$ is a recording and $t_n$ is a one-hot vector denoting to which song (or class) the recording belongs. Different versions of the same song are viewed as the samples from the same class, and different songs are regarded as the different classes. We aim to train a classification network model parameterized as $\{\theta,\lambda\}$ from $D$. As shown in \figref{fig:net}, $\theta$ is the parameter of all convolutions and FC$0$; $\lambda$ is the parameter of FC$1$; $f_\theta$ is the output of the FC$0$ layer. Then, this model could be used for cover song retrieval. More specifically, after the training, given a query $Q$ and references ${R_n}$ in the dataset, we extract latent features $f_\theta(Q),  {f_\theta(R_n)}$ which we call as music representations obtained by the network, and then we use a metric $s$ to measure the similarity between them. In the following sections, we will discuss the low-level representation used, the design of network structure and a robust trained model against key transposition and tempo change. We use lowercase for audio and uppercase for the CQT to discriminate.

\subsection{Low-level Representation}
The CQT, mapping frequency energy into musical notes, is extracted by \textit{Librosa} \cite{mcfee2015librosa} for our experiment. The audio is resampled to $22050$ Hz, the number of bins per octave is set as $12$ and Hann window is used for extraction with a hop size of $512$. Finally, a $20$-point mean down-sampling in the time direction is applied to the CQT, and the resulting feature is a sequence with a feature rate of about $2$ Hz. It could also be viewed as an $84 \times T$ matrix where $T$ depends on the duration of input audio.

\subsection{Network Structure}

\begin{figure}[h]
    \centering
    \includegraphics[trim = 2mm 2mm 2mm 2mm , clip, width=2.8in]{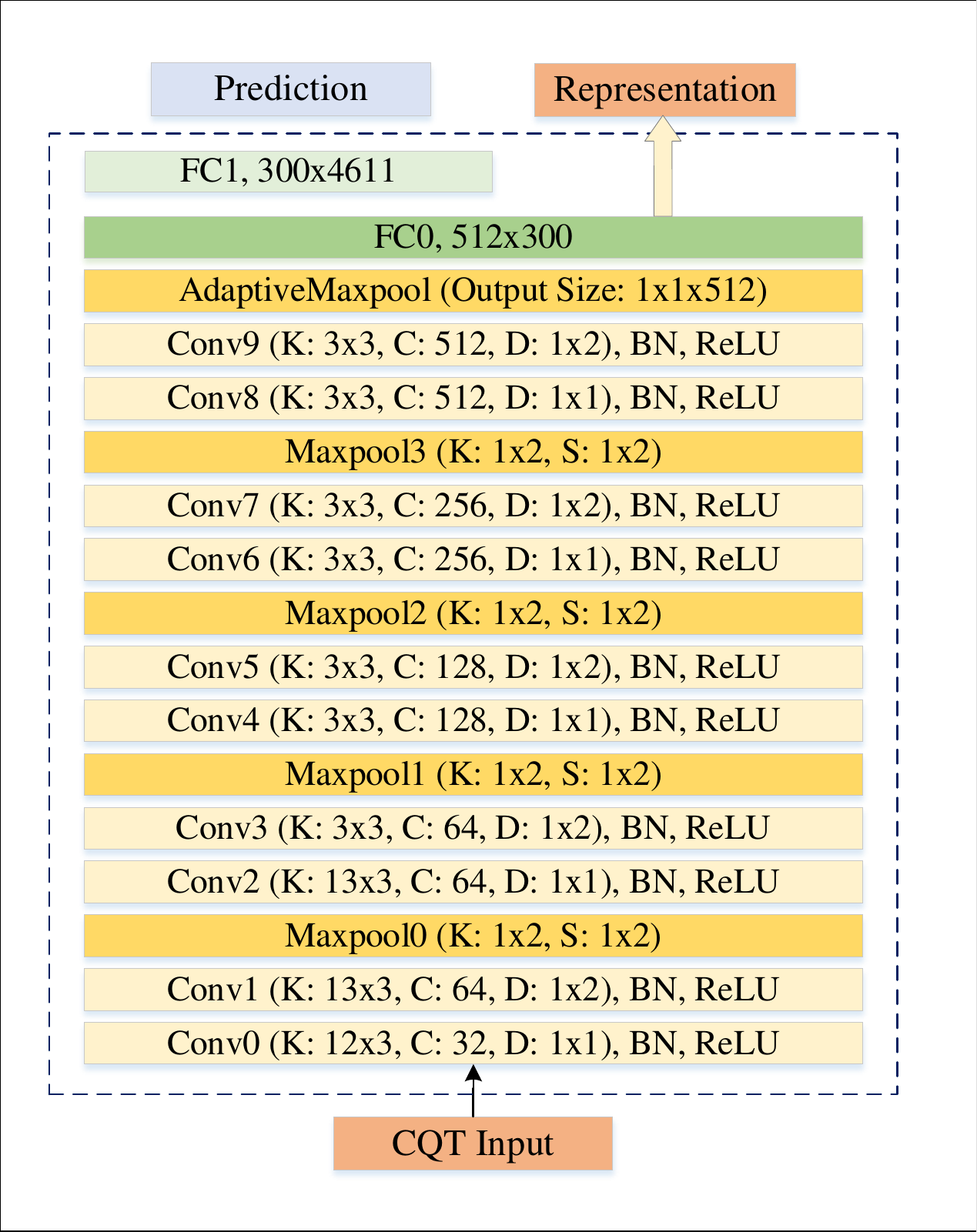}
    \caption{Network structure. K: kernel size, C: channel number, D: dilation and S: stride. The stride is set to $1\times1$ for the convolutional layers, and pooling layers has a dilation of $1\times1$. The output dimension is $4611$, the number of classes in the training set.}
    \label{fig:net}
\end{figure}

Inspired by successful applications of network architectures like \cite{simonyan2014very,he2016deep}, we design a novel Network architecture for the cover song task. We stack small filters following with max pooling operations, except that in initial layers, we design the height of filter to be $12$ or $13$ (see \figref{fig:net}) as the number of bins per octave is set as $12$ in the CQT.
This setting results in that the units of the third layer have a receptive field with a height of $36$; it spans across three octaves or thirty-six semitones. 

We also introduce dilated convolution into the model to enlarge the receptive field because cover song identification focuses on the long-term melody of the music. The design is consistent with the ideas of existing works in \cite{serra2008chroma,bertin2012large}, which extracted features or measured the similarity from a long range. 

More importantly, our model does not involve any downsample pooling operation in the frequency dimension; in other words, the vertical stride is always set to $1$, different from prevalent network structures like VGG and ResNet \cite{simonyan2014very,he2016deep}. The motivation behind this design focuses on the fact that key transposition may be one or two semitones, corresponding to moving the CQT matrix vertically for merely one or two elements. Without downsampling feature map, the network remains a higher resolution and deals with key transposition better. We experimentally validate that this design helps improve the precision (see Section \ref{subsec:expl}). 

Furthermore, after several convolutional and pooling layers, we apply an adaptive pooling layer to the feature map, whose length varies depending on the input audio. Obviously, this global pooling has the advantage of converting variable-length feature maps into fixed-dimensional vectors, connected with two fully-connected layers to exploit more information. Without the global pooling, the fully-connected layers only allow fixed-dimensional inputs, which are not the cases of music as different compositions may have different durations. Comparing with \cite{yu2019temporal} who utilized Temporal Pyramid Pooling, we utilize global pooling because it performs the same as TPP in our model. One explanation is that our convolutional network is much deeper.

Given the inputs of network $X$, the output of FC$0$ is  $f_\theta(X)\in R^{300}$, and the prediction of network is $y = \mathrm{softmax}(\lambda f_\theta(X))$. Cross-entropy loss $L$ is used for training. In our cases, different versions of a song are considered as the same category, and different songs are viewed as different classes.

\subsection{Training Scheme}

\begin{algorithm}[ht]
  \caption{Data augmentation and training strategy}
  \label{alg:alg}
  \textbf{Input}: Training set $D$, batch size $n$, changing range (a, b) \\
  \textbf{Output}: Optimized parameter $\theta$, $W$
  \begin{algorithmic}[1]
  \REPEAT
  \FOR{$L \in \{200, 300, 400$\}}
  \STATE sample a batch of recordings $B$ from the training set $D$
  \STATE $S \gets \emptyset$
  \FOR{$x \in B$}
  \STATE $r \gets$ sample from $U(a, b)$
  \STATE $x \gets$ simulate tempo changes on $x$ with a changing factor $r$
  \STATE $X \gets$ extract the CQT from $x$
  \STATE $X \gets$ crop a subsequence from $X$ with a length $l$
  \STATE $S \gets S \cup X$
  \ENDFOR
  \STATE Feed-forward with $T$
  \STATE Backpropagation to update $\theta$ and $W$
  \ENDFOR
  \UNTIL{Network converges}

  \end{algorithmic}
\end{algorithm}

For each batch, we sample some recordings from the training set and extract CQTs from them. For each CQT, we randomly crop three subsequences with a length of $200, 300$ and $400$ for training, corresponding to roughly $100$s, $150$s and $200$s, respectively. 

Despite the training set contains covers performed at different speeds, each song merely owns several covers on average for training. Moreover, as our model does not explicitly handle tempo changes in cover songs, it may be difficult for the model to learn a representation robust against tempo changes automatically. Therefore, we perform data augmentation during model training. As shown in \algref{alg:alg}, we sample a changing factor from $(0.7, 1.3)$ for each recording in the batch following uniform distribution and simulate tempo changes using \textit{Librosa} \cite{mcfee2015librosa} on the recording before cropping subsequences. 

\subsection{Retrieval}
After the training, the network is used to extract music representations. As shown in \figref{fig:pipe}, given a query $q$ and a reference $r$, we first extract their CQT descriptors $Q$ and $R$ respectively, which are fed into the network to obtain music representations $f_\theta(Q)$ and $f_\theta(R))$, and then the similarity $s$, defined as their cosine similarity, are measured.
After computing the pair-wise similarity between the query and references in the dataset, a ranking list is returned for evaluation.

\section{Experimental Settings}
\label{sec:evaluation}

\subsection{Dataset}

\textit{Second Hand Songs 100K (SHS100K)}, which is collected from \textit{Second Hand Songs website} by \cite{xu2018key}, consisting of $8858$ songs with various covers and $108523$ recordings. This dataset is split into three subsets -- \textit{SHS100K-TRAIN}, \textit{SHS100K-VAL} and \textit{SHS100K-TEST} with a ratio of $8:1:1$.

\textit{Youtube} is collected from the YouTube website, containing $50$ compositions of multiple genres \cite{silva2015music}. Each song in \textit{Youtube} has $7$ versions, with $2$ original versions and $5$ different versions and thus results in $350$ recordings in total. In our experiment, we use the $100$ original versions as references and the others as queries following the same as \cite{silva2016simple,yu2019temporal,xu2018key}.

\textit{Covers80} is a widely used benchmark dataset in the literature. It has $80$ songs, with $2$ covers for each song, and has $160$ recordings in total. To compare with existing methods, we compute the similarity of any pair of recordings.

\textit{Mazurkas} is a classical music collection consisting of 2914 recordings of 49 Chopin's Mazurkas, originated from the Mazurka Project \footnote{\url{www.mazurka.org.uk}}. The number of covers for each piece varies between $41$ and $95$. For this dataset, we follow the experimental setting of \cite{silva2016simple}.

\subsection{Evaluation}
For evaluation, we calculate the common evaluation metrics: mean average precision (MAP), precision at $10$ (P@10) and the mean rank of the first correctly identified cover (MR1). These metrics are the ones used in Mirex Audio Cover Song Identification contest \footnote{\url{https://www.music-ir.org/mirex/wiki/2019:Audio_Cover_Song_Identification}}. Additionally, query time is recorded for efficiency evaluation. All the experiments are run in a Linux server with two TITAN X (Pascal) GPUs.

\section{Experimental Result and Analysis}
\subsection{Exploration of Network Structure}
\label{subsec:expl}
Firstly we explore the kernel size of CNNs and replace the kernel size of the initial three layers with {$3\times3$}, {$7\times3$}, {$15\times3$}, {$7\times7$}, {$12\times12$}. The result of experiment shows that the height of filter to be $12$ or $13$ performs the best.

Additional, we change the vertical strides of max-pooling layers and conduct several experiments to explore its influence on accuracy. The original model is denoted as CQT-Net, and the modified network is denoted as CQT-Net$\{4\}$, CQT-Net$\{3, 4\}$ and CQT-Net$\{2, 3, 4\}$ respectively, where the numbers indicate the shape of corresponding pooling layers are changed to $(2, 2)$. For instance, CQT-Net$\{2\}$ means that Pool$2$ is replaced with a pooling operation with a stride and size of $(2, 2)$. That is, the total vertical strides for CQT-Net$\{4\}$, CQT-Net$\{3, 4\}$ and CQT-Net$\{2, 3, 4\}$ are $2, 4$ and $8$, respectively. When the vertical stride increases, MAP degrades on the four datasets consistently, as well as MR1 and P@10. 
We suppose this is because the key transposition may shift one or two semitones; the network having a higher resolution of feature dimension (that is, setting vertical stride to be $1$) could capture these changes and help improve the precision. 

 \begin{table}[H]
  \small
  \centering
  \label{tab:com}
  \begin{tabular}{cccccc}
      \toprule
      & MAP & P@10 & MR1 & Time  \\ 
      \midrule
      &\multicolumn{3}{c}{Results on \textit{Youtube}} \\
      \hline
      DPLA \cite{serra2008chroma} & 0.525 & 0.132 & 9.43 & 2420s  \\ 
      SiMPle \cite{silva2016simple} & 0.591 & 0.140 & 7.91 & 18.7s \\ 
      Fingerprinting \cite{Seetharaman2017CoverSI} & 0.648 & 0.145 & 8.27 & -  \\
      SuCo-DTW \cite{silva2018summarizing}& 0.800 & 0.180 & 3.42 & 4.59s \\
      Ki-CNN \cite{xu2018key}& 0.656 & 0.155 & 6.26 & 0.35ms \\
      TPPNet \cite{yu2019temporal}& 0.859 & 0.188 & 2.85 & 0.04ms \\
      CQT-Net & \textbf{0.917} & \textbf{0.192} & \textbf{2.50} & \textbf{0.04ms} \\
      \hline
      &\multicolumn{3}{c}{Results on \textit{Covers80}} \\
      \hline
      NCP-WIDI \cite{cheng2017effective} &0.645 & - & - &-\\
      CRP \cite{serra2009cross} & 0.544 & 0.061 & - &-\\
      Fusing \cite{chen2018fusing} & 0.625 & 0.071 & - & - \\
      Ki-CNN \cite{xu2018key}& 0.506 & 0.068 & 16.4 & 0.55ms \\
      TPPNet \cite{yu2019temporal}& 0.744 & 0.086 & 6.88 & 0.06ms \\
      CQT-Net & \textbf{0.840} & \textbf{0.091} & \textbf{3.85} & \textbf{0.06ms} \\ 
      \hline
      &\multicolumn{3}{c}{Results on \textit{Mazurkas}} \\
      \hline
      DTW \cite{silva2016simple} & 0.882 & 0.949 & 4.05 & - \\
      NCD \cite{bello2011measuring} & 0.767 & - & - & - \\
      Compression \cite{silva2013video} & 0.795 & - & - & - \\
      Fingerprinting \cite{grosche2012structure} & 0.819 & - & - & - \\
      SiMPle \cite{silva2016simple} & 0.880 & 0.952 & 2.33 & - \\
      SuCo-repeat \cite{silva2018summarizing} & 0.850 & 0.940 & 2.77 & - \\
      2DFM \cite{bertin2012large} & 0.363 & 0.578 & 15.6 & 4.76ms  \\
      Ki-CNN \cite{xu2018key}& 0.707 & 0.892 & 4.01 & 5.34ms \\ 
      CQT-Net & \textbf{0.933} & \textbf{0.956} & \textbf{2.87} & \textbf{0.50ms} \\
      \hline
      &\multicolumn{3}{c}{Results on \textit{SHS100K-TEST}} \\
      \hline
      2DFM \cite{bertin2012large} & 0.104 & 0.113 & 415 & 13.9ms  \\
      Ki-CNN \cite{xu2018key}& 0.219 & 0.204 & 174 & 21.0ms \\ 
      TPPNet \cite{yu2019temporal} & 0.465 & 0.357 & 72.2 & 3.68ms \\
      CQT-Net & \textbf{0.655} & \textbf{0.456} & \textbf{54.9} & \textbf{3.68ms} \\ 
      \bottomrule
  \end{tabular}
  \caption{Performance on different datasets (- indicates the results are not shown in original works).}
  \end{table}

\subsection{Comparison}
We compare with other state-of-the-art methods on different datasets. \tabref{tab:com} shows our approach outperforms state-of-the-art methods on all datasets. The advantages of our approach lie in without relying on complicated hand-crafted features and elaborately-designed alignment algorithms, our approach exploits massive data and feature learning and obtains high precision. By collecting a larger dataset, our approach may obtain higher precision. It is worth noting that our training set \textit{SHS100K-TRAIN} mainly consists of pop music while \textit{Mazurkas} contains classical music. Our approach outperforms state-of-the-art methods on this dataset, which indicates a good generalization ability. We do not show the result of \cite{grosche2012structure} in Mazurka Project because our test sets are different. As for the large dataset \textit{SHS100K-TEST}, our method performs much better than state-of-the-art methods.

Moreover, the query time shown in the table does not include the time of feature extracting. Therefore, our method has the same time consumed as \cite{yu2019temporal}. It extracts a fixed-dimensional feature whatever the duration of input audio is. Theoretically, it has linear time complexity, faster than sequence alignment methods with quadratic time complexity. One could find that the query time of our approach is shorter than approaches such as DPLA, SiMPle by several magnitudes. For Ki-CNN and TPPNet, they model music with a fixed-dimensional vector and have similar time complexity. In our implementation, our approach learns a $300$-dimensional representation, which is the same as TPPNet, explaining why the time consumption of our approach is the same as TPPNet.

\subsection{Result Demonstration and Error Analysis}
We listen to the Top10 retrieval results and attempt to make some analysis on \textit{SHS100K-TEST}. Our approach could identify versions when performed by different genders, accompanied by different instruments, sung in different languages, etc. Especially, as our training goal is to classify the song and different versions of the same song often have similar styles, melodic and chord structures, we find that even though some candidates in the Top10 may not be the cover of the query, but they have similar properties such as accompaniment and genre with the query. For instance, \textit{Everybody Knows This Is Nowhere} by the Bluebeaters has a similar accompaniment with that of \textit{Waiting in Vain} by Bob Marley \& The Wailers. In this sense, our approach may also be used to retrieve similar music of the query and extended to content-based music recommendation.

Furthermore, we find that Top1 precision of our model is $0.81$, suggesting that it could find a cover as the Top1 candidate for $81\%$ queries. However, it works worse in some cases. This may explain why our approach obtains a MAP of $0.655$ while only a MR1 of $54.9$ on this dataset. Most importantly, the high Top1 precision and the fast retrieval speed make our method possible to handle the real-life cover song task instead of just staying in the lab stage. 

\section{Conclusion}
Different from conventional techniques, we propose CNNs for feature learning towards cover song identification. Utilizing specific kernels and dilated convolutions to extend the receptive field, we show that it could be used to capture melodic structures underlying the music and learn key-invariant representations. By casting the problem into a classification task, we train a model that is used for music version identification. Additionally, we design a training strategy to enhance the model's robustness against tempo changes and to deal with inputs with different lengths. Combined with these techniques, our approach outperforms state-of-the-art methods on all public datasets with low time complexity. Furthermore, we show that this model could retrieve various music versions and discover similar music. Eventually, we believe our method is competent to solve real-life cover song problem.
\vfill\pagebreak

\bibliographystyle{IEEEbib}
\bibliography{strings,refs}

\end{document}